\documentclass[showpacs,twocolumn,pra]{revtex4}%
\usepackage{epsfig}
\usepackage{amsmath}
\usepackage{graphicx}
\usepackage{bm}
\usepackage{amsfonts}
\usepackage{amssymb}%
\begin{document}
\title{Theory of free space coupling to high-Q whispering gallery modes}
\author{Chang-Ling Zou$^{1}$}
\author{Fang-Jie Shu$^{2}$}
\author{Fang-Wen Sun$^{1}$}
\email{fwsun@ustc.edu.cn}
\author{Zhao-Jun Gong$^{1}$}
\author{Zheng-Fu Han$^{1}$}
\author{Guang-Can Guo$^{1}$}
\affiliation{$^{1}$Key Lab of Quantum Information, University of Science and Technology of
China, Hefei 230026 }
\affiliation{$^{2}$Department of Physics, Shangqiu Normal University, Shangqiu 476000}
\date{\today }

\begin{abstract}
A theoretical study of free space coupling to high-Q whispering gallery modes
both in circular and deformed microcavities are presented. In the case of a
circular cavity, both analytical solutions and asymptotic formulas are
derived. The coupling efficiencies at different coupling regimes for cylinder
incoming wave are discussed, and the maximum efficiency is estimated for the
practical Gaussian beam excitation. In the case of a deformed cavity, the
coupling efficiency can be higher if the excitation beam can match the
intrinsic emission well and the radiation loss can be tuned by adjusting the
degree of deformation. Employing an abstract model of slightly deformed
cavity, we found that the asymmetric and peak like line shapes instead of the
Lorentz-shape dip are universal in transmission spectra due to multi-mode
interference, and the coupling efficiency can not be estimated from the
absolute depth of the dip. Our results provide guidelines for free space
coupling in experiments, suggesting that the high-Q ARCs can be efficiently
excited through free space which will stimulate further experiments and
applications of WGMs based on free space coupling.

\end{abstract}

\maketitle

\section{Introduction}

Whispering gallery modes (WGMs) were first explained by Lord Rayleigh
\cite{Rayleigh} in the case of sound propagation in St Paul's Cathedral circa
in 1878. WGMs also exist in light waves which are almost perfectly guided
round by optical total internal reflection in low loss resonators
\cite{Richtmyer}. High quality (Q) factor excessing $10^{10}$ has been
achieved. Combining the ultra-small mode volume with the high-Q factor, the
light-matter interaction can be greatly enhanced in WGM microcavities.
Therefore, optical WGMs are gaining growing attentions in a wide range of
fields including ultra-sensitive sensors \cite{sensor}, low threshold lasers
\cite{laser}, frequency combs \cite{comb}, cavity quantum electrodynamics
(QED) \cite{cqed}, and quantum optomechanics \cite{om1,om2}. The WGMs also
affect the scattering on spherical particles \cite{Mie}, which was found to
play an important role in the glory phenomenon \cite{glory}. In the last
twenty years, the high-Q optical WGMs have been reported in various
microcavities, such as liquid droplet \cite{droplet}, microsphere
\cite{sphere}, microdisk \cite{disk}, microcylinder \cite{cylinder} and
microtoroid \cite{toroid}.

The excitation and collection of the WGMs are essential issues in practical
applications. It is taken for granted that free space coupling to WGMs is very
inefficient since the radiation loss of circular shaped microcavity is
isotropic and low, which brings the difficulty of energy transference between
outside and WGMs. Near field couplers, such as prism \cite{prism}, fiber
taper\cite{taper1,taper2} and perpendicular waveguide \cite{shuPC}, enable
high efficient excitations and collections to microresonators. Therefore,
these near field couplers are widely adapted in experiments. The underlying
mechanism of coupling process between WGMs and couplers have been well
understood\cite{CMT1,CMT2,CMT3} and many phenomena in the spectrum are
reported, such as the analogue of electromagnetically induced transparent
(EIT) \cite{EIT1,EIT2,EIT3}, asymmetric Fano line shape
\cite{shuPC,Fano1,Fano2,Fano3} and ringing phenomena \cite{ringing}. However,
there are some limitations in the near field couplers, such as large
footprint, stabilization requirement, and power limit (high excitation energy
would cause strong nonlinear effects in fiber tapers).

On another hand, the asymmetric resonant cavities (ARCs) have been
demonstrated to be able to support high-Q WGMs and give highly directional
emission \cite{ARC1,ARC2,ARC3,ARC4,ARC5,ARC6,ARC7,ARC8,ARC9}. The underlying
principle and interesting chaotic ray dynamics in such an open billiard have
been studied extensively \cite{ray1,ray2,ray3,ray4,ray5}. Especially, the
unidirectional emission cavities have been successfully designed
\cite{uni1,uni2,uni3,uni4,uni5,uni6,uni7,uni8}, which enable high efficient
collection of high-Q WGMs through free space. As the reversal of emission, the
focused beam in free space can excite the WGMs, which have been studied in
experiments \cite{fsc1,fsc2,fsc3,fsc4,fsc5,fsc6}. However, the basic features
and potentials of this free space beam coupling strategy are not fully explored.

In this paper, we present a theoretical study on free space coupling to high-Q
WGMs in both circular and deformed microcavities. First of all, the cylinder
waves incident to a circular cavity is solved analytically and asymptotically
in the cylinder coordinates, and the analytical equations are well consistent
with the standard input-output formulation. Then, we consider the excitation
of the circular cavity by Gaussian beam for practical applications, the
maximal efficiency of about $20\%$ can be achieved under phase matching
condition. After that, the excitation of deformed cavity by Gaussian beam is
studied by an abstract model of multimode interaction. The result shows that
the ARCs not only gives good matching parameters, but also can balance
radiation and non-radiation losses, which can be well adapted in experiments
for high efficient free space coupling. One interesting result is that the
asymmetric spectra is universal in the free space coupling as a result of
multimode interference, and the energy transferring can not be deduced from
the dip depth in the transmission spectrum.

\section{Circular Cavity}

\subsection{Cylinder wave}

We start with a two-dimensional circular shaped microcavity, where
electromagnetic field ($\psi$) can be solved in the cylinder coordinates
($r,\phi$) analytically. Any electromagnetic field can be decomposed in the
basis of Bessel functions $J_{m}(kr)$ and Hankel functions $H_{m}%
^{(1(2))}(kr)$ with integral $m\in(-\infty,\infty)$. Since the field intensity
is finite everywhere including the origin, thus the field inside the cavity
should be represented by ${\sum}a_{m}J_{m}(nkr)e^{im\phi}\ $when $r<r_{c}$
with $r_{c}$ is the radius of the cavity. The field outside ($r>r_{c}$) is
represented by ${\sum}c_{m}H_{m}^{(2)}(kr)e^{im\phi}$ and ${\sum}b_{m}%
H_{m}^{(1)}(kr)e^{im\phi}$, which correspond to the inward and outward
traveling cylinder waves, respectively.

For $m$-th cylinder wave $c_{m}H_{m}^{(2)}(kr)e^{im\phi}$ incident to the
cavity, only $m$-th components of cavity field and reflected outgoing wave are
nonzero, due to the conservation of angular-momentum in this axial symmetry
system. Applying the boundary conditions, the coefficients of cavity field and
outgoing field can be solved analytically with%
\begin{align}
a_{m}  &  =\frac{H_{m}^{(1)}(z)H_{m}^{(2)\prime}(z)-H_{m}^{(1)\prime}%
(z)H_{m}^{(2)}(z)}{n^{p}J_{m}^{\prime}(nz)H_{m}^{(1)}(z)-J_{m}(nz)H_{m}%
^{(1)\prime}(z)}c_{m},\label{am}\\
b_{m}  &  =-\frac{n^{p}J_{m}^{\prime}(nz)H_{m}^{(2)}(z)-J_{m}(nz)H_{m}%
^{(2)\prime}(z)}{n^{p}J_{m}^{\prime}(nz)H_{m}^{(1)}(z)-J_{m}(nz)H_{m}%
^{(1)\prime}(z)}c_{m}, \label{bm}%
\end{align}
where $z=kr_{c}$, and $p=-1(1)$ for transverse electric (TE) field (transverse
magnetic (TM) field).

\begin{figure}[ptb]
\centerline{ \includegraphics[width=0.5\textwidth]{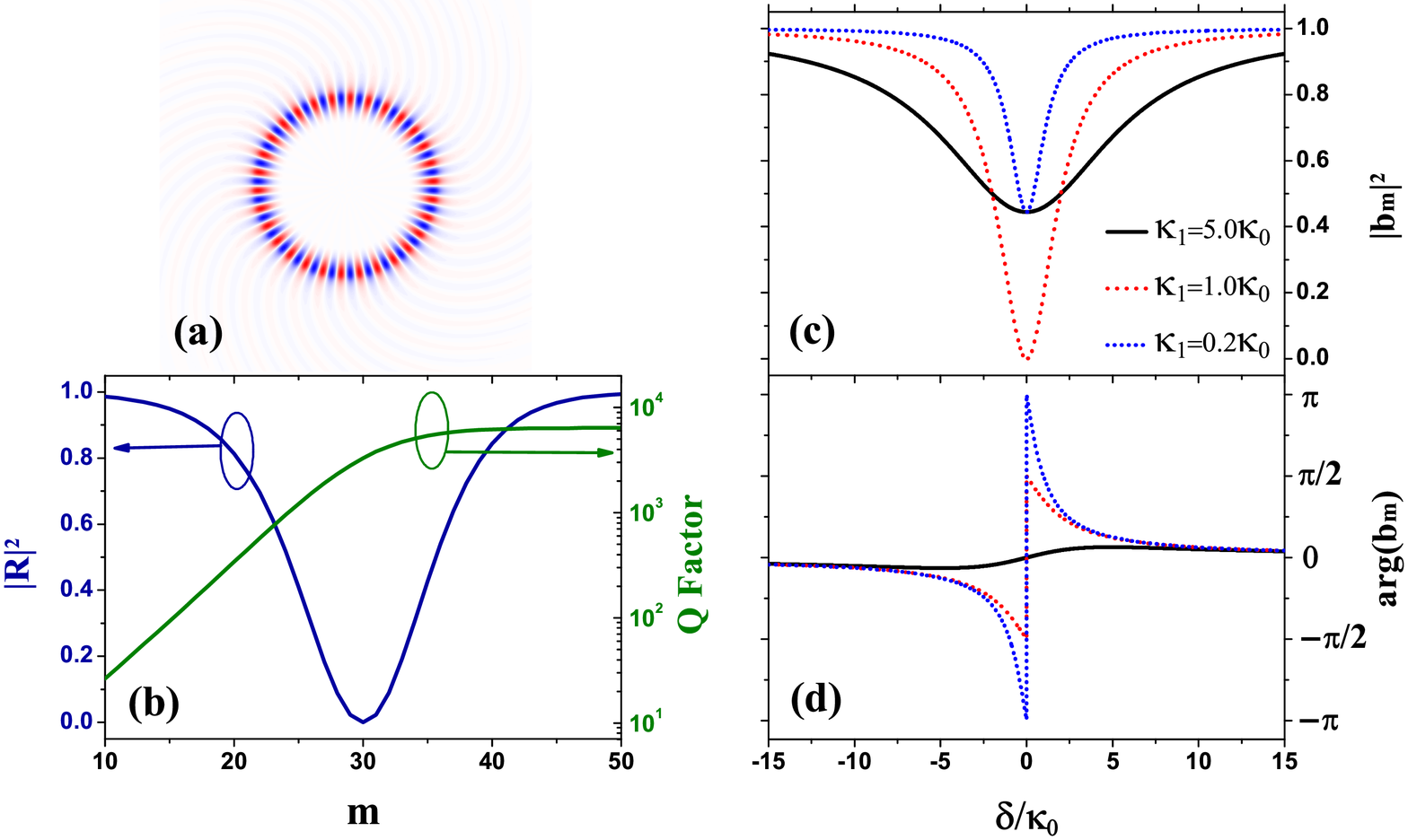}}\caption{(color
online) (a) The electric field distribution when a cylinder cavity is critical
coupling with $m=30$ cylinder wave, and detuning $\delta$ $=0$. (b) The
normalized reflectivity and the total Q factor against the angular number $m$,
with the refractive index $n=1.45+0.000117i$. (c) and (d) are the intensity
and phase of outgoing field for different coupling condition, with $\kappa
_{1}/\kappa_{0}=0.2$ (Red Solid lines), $1.0$ (Blue Dashed lines) and $5.0$
(Black Dotted lines).}%
\end{figure}

In above formulas, there are singularities at $z=z_{0}$ where
\begin{equation}
n^{p}J_{m}^{\prime}(nz_{0})H_{m}^{(1)}(z_{0})-J_{m}(nz_{0})H_{m}^{(1)\prime
}(z_{0})=0. \label{characterize}%
\end{equation}
This corresponds to the quasi-bound eigenmodes (WGMs) with dimensionless
eigenfrequency ($kr_{c}=z_{0}$), which is a complex number. When the material
refractive index $n$ is real, i.e. there no material absorption, we can write
$z_{0}=z_{0}^{r}-i\kappa_{0}$ with $z_{0}^{r}$ and $\kappa_{0}$ are real
numbers. Here, $\kappa_{0}$ is the pure radiation loss of WGM, which
corresponds to the width of the resonance in spectrum as $\kappa_{0}=z_{0}%
^{r}/2Q_{0}$ with $Q_{0}$ is the radiation quality factor.

For nearly resonant frequency $z=z_{0}+\Delta$ ($|\Delta|\ll|z_{0}|)$, by
expanding Eq. (\ref{characterize}) to the first order of $\Delta$, we can
approximately obtain
\begin{equation}
n^{p}J_{m}^{\prime}(nz)H_{m}^{(1)}(kz)-J_{m}(nz)H_{m}^{(1)\prime}(z)\approx
F(z_{0})\Delta\text{,} \label{approx0}%
\end{equation}
with $F(z_{0})=(n^{p-1}-1)[(nJ_{m}^{\prime}(nz_{0})H_{m}^{(1)\prime}%
(z_{0})+\frac{m(m-1)}{z_{0}^{2}}J_{m}(nz_{0})H_{m}^{(1)}(z_{0})]-(n^{p+1}%
-1)J_{m}(nz_{0})H_{m}^{(1)}(z_{0})$.

For the real materials, the non-radiation linear loss should be taken into
account. For example, the material absorption can be considered in the
characterize equation by adding a small complex number ($n_{i}\ll n$) to the
refractive index as $\widetilde{n}=n+in_{i}$. For a monochromatic light with
$z=z_{0}^{r}-\delta$, we have $\widetilde{n}z=(n+in_{i})(z_{0}^{r}%
-\delta)\approx n(z_{0}-\delta-i\kappa_{0}-i\kappa_{1})$ with $\kappa
_{1}=z_{0}n_{i}/n$ denotes the non-radiation loss. Substitute $\Delta
=i(i\delta+\kappa_{0}+\kappa_{1})$ into Eq. (\ref{approx0}), Eq. (\ref{am}),
and Eq. (\ref{bm}). Under the high-Q mode ($\kappa_{0}\ll z_{0}^{r}$) and near
resonance ($\delta\ll z_{0}^{r}$) conditions, and with the approximation that
most of WGM energy is confined inside the cavity, we can get%
\begin{align}
a_{m}(\delta)  &  =\frac{1}{i\delta+\kappa_{0}+\kappa_{1}}\frac{-4}{\pi
z_{0}^{r}F(z_{0}^{r})}c_{m}\text{,}\label{eq:5}\\
b_{m}(\delta)  &  =-\frac{i\delta-\kappa_{0}+\kappa_{1}}{i\delta+\kappa
_{0}+\kappa_{1}}\frac{F^{\ast}(z_{0}^{r})}{F(z_{0}^{r})}c_{m}\text{.}
\label{eq:6}%
\end{align}

On the other hand, the standard input-output formulation for a cavity mode is
\cite{Walls}
\begin{equation}
\frac{d}{dt}E_{m}=(-i\delta-\kappa_{0}-\kappa_{1})E_{m}+\sqrt{2\kappa_{0}%
}E_{m}^{in}\text{,}%
\end{equation}
where $E_{m}$ is the cavity mode field, $E_{m}^{in}$ is the incoming field in
the form of radial inward cylinder waves $H_{m}^{(2)}(z_{0})$, and the output
field is $E_{m}^{out}=-E_{m}^{in}+\sqrt{2\kappa_{0}}E_{m}.$ When the system is
in the steady state, $\frac{d}{dt}E_{m}=0$ should be satisfied, thus
\begin{equation}
E_{m}=\frac{\sqrt{2\kappa_{0}}}{i\delta+\kappa_{0}+\kappa_{1}}E_{m}%
^{in}\text{,} \label{intra}%
\end{equation}
and the output field is
\begin{equation}
E_{m}^{out}=-\frac{i\delta-\kappa_{0}+\kappa_{1}}{i\delta+\kappa_{0}%
+\kappa_{1}}E_{m}^{in}\text{.}%
\end{equation}
Comparing those equations with Eq. (\ref{eq:5}) and Eq. (\ref{eq:6}), the
cavity and outgoing fields deduced by input-output formulation are consistent
with the results derived by boundary conditions, only different in constants.
The conversion relationships between two sets of formulas are $a_{m}=\frac
{1}{\sqrt{2\kappa_{0}}}\frac{-4}{\pi z_{0}^{r}F^{\ast}(z_{0}^{r})}E_{m}$,
$b_{m}=E_{m}^{out}$, and $c_{m}=E_{m}^{in}\frac{F(z_{0}^{r})}{F^{\ast}%
(z_{0}^{r})}$.

For the case of the near field coupler, the cavity field reads \cite{CMT3}%
\begin{equation}
E_{m}=\frac{\sqrt{2\kappa_{e}}}{i\delta+\kappa_{i}+\kappa_{e}}E_{m}^{in},
\label{coupler}%
\end{equation}
where $\kappa_{i}$ is intrinsic loss including radiation and absorption
losses, and $\kappa_{e}$ is external loss induced by coupler. Comparing with
Eq. (\ref{intra}), the free space and near field coupling manners are similar,
with $\kappa_{0}$ and $\kappa_{1}$ replaced by $\kappa_{i}$ and $\kappa_{e}$.

From Eq. (\ref{intra}), the extremum of cavity field emerges when $\partial
E/\partial\kappa_{0}=0$, i.e. $\kappa_{0}=\kappa$ with $\delta=0$. Similar to
the case of a near field waveguide coupling to WGMs, there are three coupling
regimes: under coupling regime $\kappa_{0}<\kappa_{1}$; critical coupling
regime $\kappa_{0}=\kappa_{1}$; and over coupling regime $\kappa_{0}%
>\kappa_{1}$. The electric field distribution in Fig. 1(a) clearly shows the
excitation of WGM at critical coupling by an inward cylinder wave
$H_{30}^{(2)}(kr)e^{i30\phi}$ with the spiral-like propagation. No reflection
wave is excited in this case, which means all energy is perfectly absorbed by
the cavity. Fig. 1(c) and (d) are the intensity and the phase of outgoing wave
against the detuning in three regimes. In the under and over coupling regime,
the energy transfer efficiency without detuning is low. Moreover, the phase of
the outgoing wave is strongly changed by the WGM in the over coupling regime
with small detuning value, while there is only small perturbation for under
coupling regime.

In the case of near field coupler, the extra coupling loss $\kappa_{e}$ can be
controlled by simply adjusting the gap between coupler and cavity, while
$\kappa_{i}$ is an intrinsic parameter that does not change. Differently, in
the free space coupling case, $\kappa_{1}$ is an intrinsic parameter
determined by material and fabrication which is almost a constant for WGMs. It
is only possible to change $\kappa_{0}$ by change the cavity size or boundary
shape. As shown in Fig. 1(b), the Q-factor {[}$Q=kr_{c}/2(\kappa_{0}%
+\kappa_{1})${]} and the outgoing intensity change against $m$ with
$n=1.45+0.000117i$. The Q-factor increases with the increasing $m$ and shows a
saturation value of about $6\times10^{3}$. The critical coupling ($\kappa
_{0}\approx\kappa_{1}$) happens when $m=30$, corresponding to a specific
cavity size ($r_{c}\approx m/nk$). For larger or smaller cavity size, the
coupling efficiency reduces.

\subsection{Gaussian Beam}

\begin{figure}[ptb]
\centerline{ \includegraphics[width=0.45\textwidth]{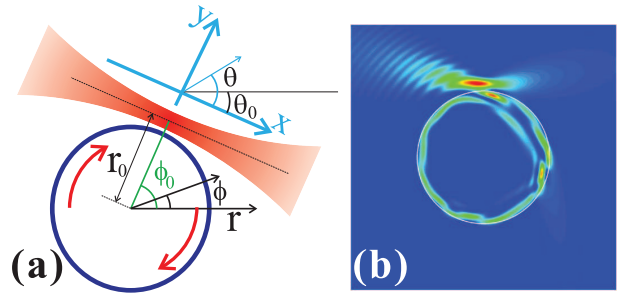}}\caption{(color
online) (a) Schematic illustration of a Gaussian beam incident to a whispering
gallery microcavity. The direction of beam is defined by $\theta_{0}$, and the
location of center is $(r_{0},\phi_{0})$ in the cylinder coordinator. (b) The
electric field intensity distribution when the incident Gaussian beam is
resonant with the WGM, with $kr_{c}=30.10,$ $r_{0}=1.1r_{c}$, $\phi_{0}=\pi/2$
and $\theta_{0}=0.$}%
\end{figure}

Gaussian beam is most widely applied in experiments. Thus we use this beam in
following studies for realistic consideration. Supposing a Gaussian beam
centered at ($x_{0},y_{0}$) with intensity $I=1$ and propagation along
$x$-axis with the waist width $w$, the electric field at the waist is
\begin{equation}
G(x_{0},y)=e^{-\frac{(y-y_{0})^{2}}{w^{2}}}%
\end{equation}
Correspondingly, the angular spectrum is
\begin{align}
\widetilde{G}(\theta)  &  =\frac{1}{2\pi}\int_{-\infty}^{\infty}%
e^{-iyk\mathrm{sin}\theta}e^{-\frac{y^{2}}{w^{2}}}dy\nonumber\\
&  =\frac{w}{2\sqrt{\pi}}e^{-\frac{k^{2}w^{2}}{4}\mathrm{sin}^{2}\theta
}\text{,}%
\end{align}
with $\theta\in(-\frac{\pi}{2},\frac{\pi}{2}).$ So, in the cylinder
coordinator in Fig. 2(a), the field distribution can be expressed as
\begin{equation}
E(r,\phi)=\int_{-k}^{k}\frac{w}{2\sqrt{\pi}}e^{-\frac{k^{2}w^{2}}%
{4}\mathrm{sin}^{2}\theta}e^{ik(\mathrm{sin}\theta y+\mathrm{cos}\theta
x)}d(k\mathrm{sin}\theta), \label{gaussian1}%
\end{equation}
where
\begin{align}
x  &  =r\mathrm{cos}(\phi+\theta_{0})-r_{0}\mathrm{cos}(\phi_{0}+\theta
_{0}),\\
y  &  =r\mathrm{sin}(\phi+\theta_{0})-r_{0}\mathrm{sin}(\phi_{0}+\theta_{0}).
\end{align}
Substituting them to the Eq. (\ref{gaussian1}) and employing the Jacobi-Anger
expansion,
\begin{equation}
e^{ikr\mathrm{cos}(\phi+\theta_{0}-\theta)}=\underset{m=-\infty}%
{\overset{\infty}{{\textstyle\sum}}}J_{m}(kr)e^{im[\pi/2-(\phi+\theta
_{0}-\theta)]},
\end{equation}
we can derive the coefficient of Gaussian beam in inward cylinder waves
expansion
\begin{equation}
c_{m}=\frac{e^{im(\pi/2-\theta_{0})-ikq-\frac{[kr_{0}d-m]^{2}}{k^{2}%
w^{2}+i2kq}}}{2\sqrt{1+i\frac{2q}{kw^{2}}}}, \label{Gaussian}%
\end{equation}
in the limit of $kw\gg1$ and $d=r_{0}\mathrm{sin}(\phi_{0}+\theta_{0})$ and
$q=r_{0}\mathrm{cos}(\phi_{0}+\theta_{0})$. Here, only the part of
$\mathrm{sin}\theta\ll1$ contributes in the integral, and the approximation is
applied
\begin{equation}
\mathrm{cos}\theta=\sqrt{1-\mathrm{sin}^{2}\theta}\approx1-\frac{1}%
{2}\mathrm{sin}^{2}\theta\text{.}%
\end{equation}

Substituting the coefficients to Eq. (1) and (2), the intracavity field and
outgoing field can be solved analytically for a cylinder cavity. For example,
Fig. 2(b) shows the near field distribution when a Gaussian beam incident to
the cavity with the on-resonance frequency to WGM with $m=40$. Obviously, most
energy are directly transmitted. Only a small portion of energy can be coupled
into the cavity field. By integrating the Poynting vector of the
electromagnetic field, we can calculate the power (energy flux) contained in
the Gaussian beam
\begin{equation}
P_{G}=\sqrt{\frac{\pi}{2}}\frac{kw}{2\omega\mu},
\end{equation}
where $\omega$ is angular frequency, and $\mu$ is relative permeability. Power
of normalized cylinder wave $H_{m}^{(1(2))}(kr)e^{im\phi}$ is
\begin{equation}
P_{m}=\frac{2}{\omega\mu}.
\end{equation}
Therefore, the ratio of power contained in the $m$-th cylinder waves to that
in the Gaussian beam is
\begin{equation}
\eta_{m}=\left\vert c_{m}\right\vert ^{2}\frac{P_{m}}{P_{G}}.
\end{equation}
From Eq. (\ref{Gaussian}), $\eta_{m}$ is maximal when $q=0$ and $d=m/k$. The
first condition guarantees the smallest expansion of cylinder waves in
momentum space ($m$). The second condition corresponds to the \emph{phase
matching condition} that the momentum of the pump beam ($kd$) should be equal
to the momentum of the WGM ($m$). In practical applications, we can manipulate
the beam position \textit{d} to satisfy the conditions, then
\begin{equation}
\eta_{max}=\sqrt{\frac{2}{\pi}}\frac{1}{kw}.
\end{equation}
Therefore, the rate of power transfer from a Gaussian beam to the cavity WGM
is limited to $\eta_{max}$. Under the paraxial approximation, the waist $w$ of
a Gaussian beam should be larger than $4/k$, i.e. $kw\geq4$, corresponding to
the energy transferring rate $\eta_{m}\leq0.2$ for a cylinder cavity.

\section{Deformed Cylinder}

\subsection{General Model}

When the cavity boundary is slightly deformed
\cite{ARC1,ARC2,ARC3,ARC4,ARC5,ARC6,ARC7,ARC8,ARC9,uni2,uni3,uni4,uni5,uni7,uni8}
or small perturbations are introduced to the cavity \cite{uni1,uni6}, highly
directional emission can be obtained. Because the boundaries of ARCs are not
regular, different momentum (m) components can couple to each other. In
addition, WGMs always exist in pairs, clockwise ($A_{c}$) and anti-clockwise
($A_{a}$) traveling modes, in a circular-like shape microcavity, then the
general Hamiltonian of ARC reads
\begin{equation}
\mathcal{H}=\mathcal{H}_{modes}+\mathcal{H}_{int}+\mathcal{H}_{pump}.
\end{equation}
Here,
\begin{equation}
\mathcal{H}_{modes}=\sum_{j}\hbar\delta_{j}(A_{j,c}^{\dag}A_{j,c}%
+A_{j,a}^{\dag}A_{j,a})
\end{equation}
represents the energy of cavity modes, with $\delta_{j}$ is the frequency
difference between cavity mode and excitation laser, and $A_{j,c}^{\dag
}(A_{j,c})$ is the creation (annihilation) operator of cavity mode. The
coupling between different modes is
\begin{align}
\mathcal{H}_{int}  &  =\sum_{j}\sum_{k>j}(g_{j,k,c}A_{j,c}^{\dag}%
A_{k,c}+g_{j,k,a}A_{j,a}^{\dag}A_{k,a}+h.c.)\nonumber\\
&  +\sum_{j,k}(\beta_{j,k}A_{j,c}^{\dag}A_{k,a}+h.c.),
\end{align}
where $\beta_{j,k}$ is the coupling coefficient between counter propagation
WGMs and $g_{j,k,a(c)}$ is the coupling coefficient between co-propagation
modes. The external pumping on modes is given by
\begin{equation}
\mathcal{H}_{pump}=\sum_{j,m}(ih_{j,m,c}A_{j,c}^{\dag}u_{m}^{in}%
+ih_{j,m,a}A_{j,a}^{\dag}u_{-m}^{in}+h.c.).
\end{equation}
Here, $h_{j,m,c}$ is the coupling strength of the cavity mode to the outside
cylinder waves, with $m=1\cdots\infty$, where the external incoming field and
outgoing field are represented by cylinder waves as ${\textstyle\sum}%
u_{m}^{in}e^{i\varphi_{m}}H_{m}^{(2)}(kr)e^{-im\phi}$ and ${\textstyle\sum
}u_{m}^{out}e^{-i\varphi_{m}}H_{m}^{(1)}(kr)e^{-im\phi}$ with $e^{-i2\varphi
_{m}}=-H_{m}^{(2)}(kr_{c})/H_{m}^{(1)}(kr_{c})$. Actually, $A_{c}$ and $A_{a}$
are time reversal to each other, so the coupling coefficients satisfy
$g_{j,k,c}=g_{j,k,a}^{\ast}=g_{j,k}$ and $h_{j,m,c}=h_{j,m,a}^{\ast}=h_{j,m}$.

Here, we only concern about the ARCs with smooth boundary and small
deformation which can support high-Q WGMs. Then we can omit the coupling
between counter propagation WGMs ($\beta_{i,j}=0$) and the direct scattering
induced transition between incoming and outgoing cylinder waves. Thus the
Heisenberg equation of cavity field are written as
\begin{align}
\frac{d}{dt}A_{j,c}  &  =-i\sum\limits_{k>j}g_{j,k}A_{k,c}-i\sum
\limits_{k<j}g_{k,j}^{\ast}A_{k,c}\nonumber\\
&  -\chi_{j}A_{j,c}+\sum\limits_{m}h_{j,m}u_{m}^{in},
\end{align}
where $\chi_{j}=i\delta_{j}+\kappa_{j,0}+\kappa_{j,1}$, $\kappa_{j,0}=\sum
_{m}\left\vert h_{j,m}\right\vert ^{2}/2$ and $\kappa_{j,1}$ are intrinsic
radiation loss and non-radiation loss, respectively. Supposing there is only
one high-Q WGM with $j=1$ near resonance of the excitation, and other modes
are low-Q or largely detuned, we can adiabatic eliminate the fast varying
terms by $\frac{d}{dt}A_{j,c}=0$ for $j\geq2$, i.e.
\begin{equation}
A_{j,c}|_{j\geq2}=-i\frac{g_{1,j}^{\ast}}{\chi_{j}}A_{1,c}+\underset
{m=1}{\overset{\infty}{{\textstyle\sum}}}\frac{h_{j,m}}{\chi_{j}}u_{m}%
^{in}\text{.}%
\end{equation}
Therefore,
\begin{equation}
\frac{d}{dt}A_{1,c}=-\widetilde{\chi}_{1}A_{1,c}+\underset{m=1}{\overset
{\infty}{{\textstyle\sum}}}\widetilde{h}_{1,m}u_{m}^{in}\text{.}%
\end{equation}
Here, $\widetilde{\chi}_{1}=\chi_{1}+\kappa_{e}+i\delta_{e}$ with $\kappa
_{e}=\sum_{j\geq2}\frac{\kappa_{j,0}+\kappa_{j,1}}{\left\vert \chi
_{j}\right\vert ^{2}}\left\vert g_{1,j}\right\vert ^{2}$ and $\delta_{e}%
=-\sum_{j\geq2}\frac{\delta_{j}}{\left\vert \chi_{j}\right\vert ^{2}%
}\left\vert g_{1,j}\right\vert ^{2}$, corresponding to the extra loss and
frequency shift induced by low-Q modes, and $\widetilde{h}_{1,m}=h_{1,m}%
-i\sum_{j\geq2}g_{1,j}\frac{h_{j,m}}{\chi_{j}}$ are effective coupling
coefficients to outside for high-Q WGM. Denoting the coupling efficiencies and
the incoming field by vectors $\vec{h}=\{\widetilde{h}_{1,m}\}$ and $\vec
{u}=\{u_{m}^{in}\}$ with $m=1\cdots\infty$, the cavity field of steady state
becomes
\begin{equation}
A_{1,c}=\xi\left\vert \vec{h}\right\vert \left\vert \vec{u}\right\vert
/\widetilde{\chi}_{1},
\end{equation}
where $\xi=\vec{h}\cdot\vec{u}/\left\vert \vec{h}\right\vert \left\vert
\vec{u}\right\vert $ is the \textit{beam matching parameter}. The output for
arbitrary incoming field reads
\begin{equation}
u_{m}^{out}=-u_{m}^{in}+f_{m}A_{1,c}+\sum_{j\geq2}\sum_{m^{\prime}}%
\frac{h_{j,m}^{\ast}h_{j,m^{\prime}}}{\chi_{j}}u_{m^{\prime}}^{in},
\label{out}%
\end{equation}
where $f_{m}=h_{1,m}^{\ast}-i\sum_{j\geq2}\frac{g_{1,j}^{\ast}}{\chi_{j}%
}h_{j,m}^{\ast}$ corresponds to the radiation of $A_{1,c}$.

From Cauchy-Schwartz inequality, $\left\vert \xi\right\vert \leq1$, the
maximum of matching can be achieved only when
\begin{equation}
\vec{u}\propto\vec{h}^{\ast}/\left\vert \vec{h}\right\vert .
\end{equation}
This corresponds to the optimal beam to excite the WGM $E_{inc}(r,\phi
)\propto\sum_{m}e_{m}^{i\varphi_{m}}\widetilde{h}_{1,m}^{\ast}H_{m}%
^{(2)}(kr)e^{-im\phi}$, which is the reverse of the radiation of
anti-clockwise WGM $E_{a}^{rad}(r,\phi)\propto\sum_{m}\widetilde{h}%
_{1,m}e^{-i\varphi_{m}}H_{m}^{(1)}(kr)e^{im\phi}$, meaning that the optimal
excitation of clockwise WGM can be achieved by the reverse of the radiation of
anti-clockwise WGM.

When on resonance, the cavity field reads $A_{1,c}=\xi\left\vert \vec
{h}\right\vert \left\vert \vec{u}\right\vert /(\kappa_{1,0}+\kappa
_{1,1}+\kappa_{e})$, and the energy transferring depends on both $\xi$ and
$\left\vert \vec{h}\right\vert /(\kappa_{1,0}+\kappa_{1,1}+\kappa_{e})$. All
the parameters mentioned above can not be solved analytically because of the
irregular boundary shape. Realistic details about the free space coupling
should be solved by numerical simulation \cite{fscshu}. In the following, we
analytically solve the free space coupling to ARCs by an abstract model
without the specific boundary shape. In this simplified model, we assume
directional emission is in the form of Gaussian beam ignoring the specific
boundary of a ARC. The general underlying mechanism and basic properties are
revealed with several reasonable approximations of a near circular boundary shape.

\begin{figure}[ptb]
\centerline{ \includegraphics[width=0.45\textwidth]{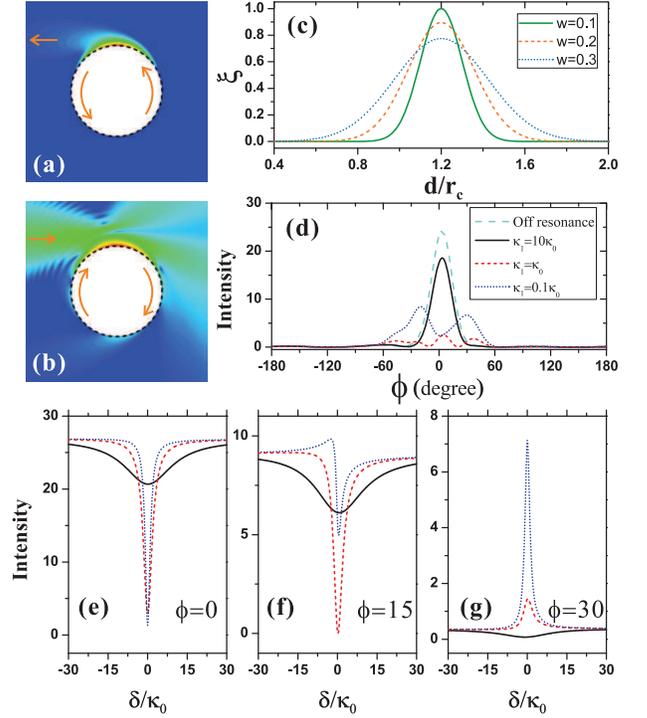}}\caption{(color
online) (a) Field distribution of directional emission of anti-clockwise WGM
in a near circular cavity (The field is shown in Logarithm scale, and the
intracavity field is not shown since the exact boundary shape is not known in
our abstract model). The emission is in the form of Gaussian beam with
$d_{t}=1.2r_{c}$, $w_{t}=0.1r_{c}$ and direction is shown by arrows. (b) Field
distribution when an on-resonance Gaussian beam coupling to the cavity in the
over coupling regime with $\kappa_{1}=0.1\kappa_{{0}}$, $d=1.2$, $w=0.2r_{c}$
and direction is shown by arrows. (c) Beam matching parameters against the $d$
for different $w$. (d) The far field intensity of outgoing wave when a
Gaussian beam incident to a cylinder, for excitation frequency off-resonance
and on-resonance at difference coupling conditions. (e), (f) and (g) are
spectra detected at different angle ($\phi$) in far field.}%
\end{figure}

\subsection{Coupling Through Barrier Tunneling}

In a very slightly deformed ARC, the coupling between high-Q and low-Q modes
are weak, and the directional emission is mainly due to the direct tunneling
\cite{Creagh}. The center of the emission beam is departed from the boundary,
as a result of barrier tunneling \cite{Tomes}. We assume that the parameters
of emission Gaussian beam are $d_{t}=1.2r_{c}$, $w_{t}=0.1r_{c}$ and direction
along $x$-axis, as shown in Fig. 3(a). Thus, the coupling coefficients can be
solved from the mode emission
\begin{equation}
h_{1,m}=\mathfrak{h}_{0}e^{-im\pi/2-\frac{[kd_{t}-m]^{2}}{k^{2}w_{t}^{2}%
}+i\varphi_{m}},
\end{equation}
where $\mathfrak{h}_{0}=\sqrt{\kappa_{0}/\sqrt{\frac{\pi}{2}}\frac{kw}{2}}$,
corresponding to the normalization ${\sum}\left\vert h_{1,m}\right\vert
^{2}=2\kappa_{0}$.

We discussed the coupling in detail using a Gaussian beam from left with
$\theta_{0}=0$ and on-resonance to a high-Q mode of $kr_{c}=31.1$. Firstly, we
calculated the beam matching parameter $\xi$ versus the beam position $d$ for
different beam widths $w$, as shown in Fig. 3(c). The result shows that good
match with $\xi>0.8$ can be easily realized with good fault tolerant to the
imperfect focus or align of beam in experiments. In the near integrated
system, we can neglect the coupling to other low-Q modes $(\kappa_{0}%
^{1,c}+\kappa_{1}^{1,c}\gg\kappa_{e})$, then $A_{c}\approx\sqrt{2\kappa_{1,0}%
}/\left(  \kappa_{1,0}+\kappa_{1,1}\right)  $. Similar to the cylinder case,
the critical coupling with $\kappa_{1,0}=\kappa_{1,1}$ guarantees the maximum
energy transfer.

With these approximations, we can calculate the far field distributions by
substituting parameters into Eq. (30). The far field distributions at
different coupling regime are shown in Fig. 3(d) with $w=0.2r_{c}$. At
critical coupling, the outgoing energy is great reduced but larger than $0$ as
a result of imperfect matching. In under coupling, there is a small reduction
of energy in all directions. However, in over coupling regime, the peak at
$\phi=0$ is split into two peaks, which is attributed to the interference
between the transmitted beam and the emission of the cavity mode, as a result
of the strong phase shift at the over coupling regime.

The spectra are also calculated (Figs. 3(e), 3(f) and 3(g)) in different far
field angles. It is revealed that the line shapes strongly depend on the
positions of detectors. At $\phi=0^{\circ}$, the resonances shows regular
Lorentz-shape dips similar to traditional near field coupler. In contrast with
Fig. 1(c), similar dip depths are given in the over and critical coupling
condition regimes, which indicates that the dip depth can not tell us the
energy transfer rate in the free space coupling. Asymmetric line shapes
(Fano-like and EIT-like) are observed at $\phi=15^{\circ}$ and $30^{\circ}$,
as a result of multiple cylinder waves interference formalized in Eq.
(\ref{out}).

\subsection{Coupling Through Refraction}

\begin{figure}[ptb]
\centerline{ \includegraphics[width=0.45\textwidth]{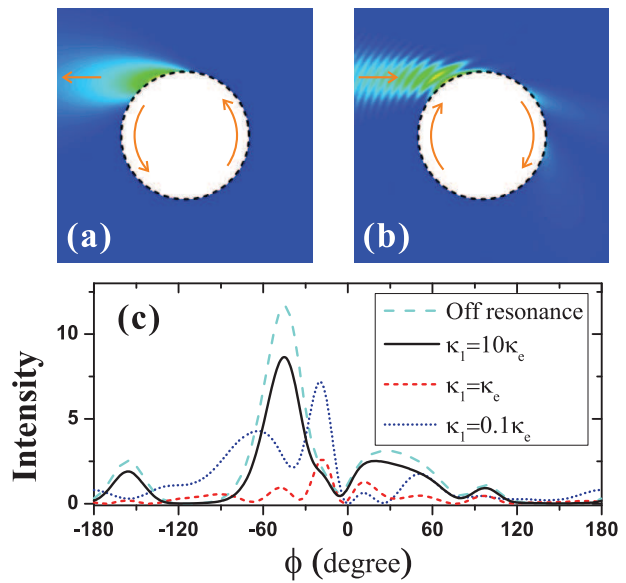}}\caption{(color
online) (a) Field distribution of directional emission of anti-clockwise WGM
in a near circular cavity (The field is shown in normal scale, and the
intracavity field is not shown since the exact boundary shape is not known in
our abstract model). The emission is in the form of Gaussian beam with
$d_{r}=0.9r_{c}$, $w_{r}=0.1r_{c}$ and direction is shown by arrows. (b) The
field distribution when an on-resonance Gaussian beam coupling to the cavity
in the over coupling regime with $\kappa_{1}=0.1\kappa_{0}$, $d=0.9$,
$w=0.2r_{c}$ and direction is shown by arrows. (c) The far field intensity of
outgoing wave when a Gaussian beam is incident to the ARC with Gaussian beam
emission, for off-resonance and on-resonance under different coupling
conditions.}%
\end{figure}

When the degree of deformation is large, the directional emission is mainly
due to the radiation from low-Q modes. This process is also known as the
dynamical tunneling in the study of ARC, where rays in high-Q mode tunnel into
chaotic sea and refracted out at some specific region in phase space
\cite{ARC5,ARC6}. The center of Gaussian beam should be adjusted on the cavity
as a result of refraction \cite{image}.
For the refraction dominated emission, the extra loss induced by low-Q\ mode
is much larger than the intrinsic radiation loss ( i.e. $\kappa_{e}\gg
\kappa_{0}$), thus we can neglect the direct tunneling loss. Assuming the
emission is Gaussian beam with the parameters $d_{r}=0.9r_{c}$, $w_{r}%
=0.1r_{c}$ and direction along $x$-axis (Fig. 4(a)), we have
\begin{equation}
\sum_{j\geq2}\frac{g_{1,j}}{\chi_{j}}h_{j,m}e^{-i\varphi_{m}}=\mathfrak{h}%
_{e}e^{-im\pi/2-\frac{[kd_{t}-m]^{2}}{k^{2}w_{t}^{2}}}.
\end{equation}
Here, $\mathfrak{h}_{e}=\sqrt{\kappa_{e}/\sqrt{\frac{\pi}{2}}\frac{kw}{2}}$
with $\kappa_{e}=\frac{1}{2}\sum_{m\geq2}\frac{\left\vert g_{1,m}\right\vert
^{2}\left\vert h_{m,m}\right\vert ^{2}}{\left\vert \chi_{m}\right\vert ^{2}}$.
In addition, the interactions between low-Q modes are not taken into
consideration. We also assume the deformation just slightly influence the
low-Q modes for simplicity, thus $h_{j,m}=0$ if $j\neq m$. Therefore
\begin{equation}
\frac{g_{1,m}}{\chi_{m}}h_{m,m}=\mathfrak{h}_{e}e^{-im\pi/2-\frac
{[kd_{t}-m]^{2}}{k^{2}w_{t}^{2}}+i\varphi_{m}}.
\end{equation}
Similar to the procedure in previous section, we solved the outgoing field
when a Gaussian beam incident from left with $\theta_{0}=0$, $w=0.2r_{c}$,
$d=0.9r_{c}$ and on-resonance to a high-Q mode with $kr_{c}=31.1$. As shown in
Figs. 4(b) and 4(c), the maximal far field intensity is located away from
$\phi=0$ as a result of the refraction of incident beam. The far field
distribution is similar to the case of tunneling coupling, and the EIT or
asymmetric line shapes are presented in the spectra (not shown here).

\section{Discussion}

According to the above analysis, the coupling to high-Q WGMs through free
space can be efficient. For the cylinder wave, the largest coupling efficiency
can be achieved under the phase matching condition, and the maximum is limited
by waist width of a focused Gaussian beam, as the ratio of power contained in
the cylinder wave is limited. For a deformed cavity, the efficiency can be
higher since the pump Gaussian beam can be well matched to the directional
emission beam.

The energy transferred to the cavity is also determined by the coupling
regime, i.e. the ratio of radiation loss to non-radiation loss. The strongest
coupling happens when $\kappa_{0}\approx\kappa_{1}$. The radiation loss
$\kappa_{0}$ decreases exponentially with the increasing of cavity size, but
the non-radiation loss $\kappa_{1}$ is almost constant. Therefore, in a large
cavity, $\kappa_{0}$ is much greater than $\kappa_{1}$, which will give rise
to very inefficient coupling. This limitation can be removed in the deformed
cavity, as $\kappa_{0}$ can be tuned by adjusting the deformation. Usually, a
large degree of deformation can also lead to better directionality of cavity
emission, thus better beam matching. Therefore, we can always tune the
deformation degree to achieve the largest energy transfer rate, similar to the
tuning of gap between near field coupler and microcavity.

In contrast to near field coupler to microcavity, the asymmetric line shapes
are universal in free space coupling to WGMs \cite{uni8,fscshu}. In more
realistic cases, maybe multiple high-Q modes near to each other in spectrum
are involved, and then the interactions between high-Q modes are enhanced by
the low-Q modes mediated mode-mode interaction. From the view of ray dynamics
in ARC, the high-Q modes usually correspond to regular period or quasi-period
orbits. Due to dynamical tunneling, the high-Q modes can couple to chaotic
modes which correspond to chaotic ray trajectories. Therefore, the chaotic sea
mediated the indirect coupling between separated regular orbits in phase space
(high-Q modes), which is also known as chaos assisted tunneling
\cite{ray3,CAT1,CAT2}. Similar to the case of waveguide coupled microcavity
\cite{EIT3,Fano1}, we can expect more profound asymmetric spectra when take
more high-Q modes into consideration.

\section{Conclusion}

In summary, we present the theoretical study of the free space coupling to
high-Q WGMs in both regular and deformed microcavities. The coupling by free
space beam depends on the \textit{phase matching} and \textit{beam matching}
conditions in regular cylinder cavities and ARCs, respectively. Three coupling
regimes of free space coupling are discussed: the cavity should work near the
critical coupling regime for high efficiency energy transfer. In cylinder
cavity, the maximum energy transfer efficient is limited, and critical
coupling can only be achieved with specific cavity size. The ARCs not only
give good beam matching between focused Gaussian beam and highly directional
emission, but also enable the tuning of the cavity degree of deformation to
achieve critical coupling. Therefore, the efficiency approaching unity is
possible in realistic experiments of free space coupling to unidirectional
emission cavity. It is found that the asymmetric spectra or peak like spectra
instead of the Lorentz-shape dip is universal in spectra, and the coupling
efficiency cannot be estimated from the absolute depth of dip. Our results
provide guidelines for free space coupling to high-Q WGMs, which will be
valuable for further experiments and applications of WGMs based on free space coupling.

\section{Acknowledgement}

We thank Prof. Hailin Wang and Prof. Kyungwon An for discussions and comments.
This work was supported by the 973 Programs (No. 2011CB921200), the National
Natural Science Foundation of China (NSFC) (No. 11004184), the Knowledge
Innovation Project of the Chinese Academy of Sciences (CAS). F.-J. Shu is
supported by the Foundation of He'nan Educational Committee (No. 2011A140021)
and the Young Scientists Fund of the National Natural Science Foundation of
China (Grant No. 11204169).

\end{document}